\documentclass{ws-mpla}
\usepackage{graphicx}
\usepackage{psfrag}
\usepackage{bm}
\usepackage{epsfig}
\usepackage{amssymb}
\begin{document}
\title{Quasiparticle universes in Bose-Einstein condensates}
\author{Uwe R. Fischer}
\address{Eberhard-Karls-Universit\"at T\"ubingen, 
Institut f\"ur Theoretische Physik \\ 
Auf der Morgenstelle 14, D-72076 T\"ubingen, Germany}

\maketitle
\pub{Received (Day Month Year)}{Revised (Day Month Year)}

\begin{abstract}
Recent developments in simulating 
fundamental quantum field theoretical effects in the kinematical 
context of analogue gravity are reviewed. Specifically, it is argued that a 
curved space-time generalization of the Unruh-Davies effect 
-- the Gibbons-Hawking effect in the 
de Sitter space-time of inflationary cosmological models 
-- can be implemented and verified in an ultracold gas of bosonic atoms.
\end{abstract}
\ccode{PACS Nos.: 04.62.+v, 03.75.Kk, 04.70.Dy, 98.80.Es}
\maketitle


\section{The concept of an effective space-time metric}
Curved space-times are familiar from Einstein's theory 
of gravitation,\cite{MTW} where the metric tensor ${\sf g}_{\mu\nu}$, 
describing distances in a curved 
space-time with local Lorentz invariance, is determined
by the solution of the Einstein equations. A major problem 
for an experimental investigation of the (kinematical as well 
as dynamical) properties of curved space-times is that 
generating a significant curvature, equivalent to a (relatively) small
curvature radius, is a close to impossible undertaking 
in manmade laboratories. 
For example, the effect of the gravitation of the whole Earth  
is to cause
a deviation from flat space-time on this planet's surface 
of only the order of $10^{-8}$
(the ratio of Schwarzschild and Earth radii). 
The fact that proper gravitational effects are intrinsically small 
is basically due to the smallness of Newton's gravitational constant
$G=6.67\times 10^{-11}$\,m$^3$kg$^{-1}$sec$^{-2}$.
Various fundamental classical and quantum 
effects in strong gravitational fields, 
a few of which we discuss below, are thus inaccessible for 
Earth-based experiments. The realm of strong
gravitational fields (or, equivalently,
rapidly accelerating a reference frame to simulate gravity
according to the equivalence principle), is therefore difficult to reach. 
However, Earth-based experiments are desirable, 
because they have the obvious advantage that they 
can be prepared and, in particular, repeated under possibly different 
conditions at will.

The formalism to be described in what follows is aimed at the realization of 
{\em effective} curved space-time geometries in perfect fluids, which 
{can} indeed be prepared on Earth, and which mimic the effects of
gravity inasmuch the {kinematical} 
properties of curved space-times are concerned. 
Among such perfect fluids are Bose-Einstein condensates, i.e., the dilute 
matter-wave-coherent gases formed 
if cooled to ultralow temperatures, where 
the critical temperatures are of order $T_c\sim 100\,{\rm nK}\cdots 1\, \mu$K; 
for a short review of the (relatively) recent status 
of this rapidly developing field see Ref.\,\cite{Anglin}. In what follows, 
it will be of some importance that 
Bose-Einstein condensates belong to a special class of {quantum}
perfect fluids, so-called superfluids.\cite{Khalatnikov}

The curved space-times we have in mind 
are experienced by sound waves propagating
on the background of a spatially and temporally inhomogeneous perfect fluid.
Of primary importance is, first of all,  
to realize that the identification of sound waves
propagating on an inhomogeneous background, which is 
itself determined by a solution of 
Euler and continuity equations, and photons 
propagating in a curved space-time, which is determined by a 
solution of the Einstein equations, is of a {\em kinematical} nature. 
That is, the  space-time metric is fixed externally by a background
field obeying the laws of hydrodynamics (which is prepared by the 
experimentalist), and not self-consistently by a solution of the Einstein
equations 
$G_{\mu\nu} = 
8\pi T_{\mu\nu} $ (where $G$ and the speed of light are set to unity). 
The latter equations 
relate space-time curvature -- represented by the Einstein tensor $G_{\mu\nu}
= R_{\mu\nu} - \frac12 g_{\mu\nu} R$, where $R_{\mu\nu}$ is the
Ricci tensor and $R= g^{\mu\nu} R_{\mu\nu}$ the Ricci scalar --  
with the energy-momentum content of all other fundamental quantum fields. 
This energy-momentum content 
is represented by the classical quantity $T_{\mu\nu}$, which is the 
regularized expectation value of a quantum energy-momentum 
tensor.\cite{BirrellDavies} 

As a first introductory step to understand the nature of the proposed 
analogy, consider the wave equation for the velocity potential 
of the sound field $\Phi$, which in a homogeneous medium at rest reads 
\begin{equation} 
\left[-\frac 1{c_s^2} \frac{\partial^2}{\partial t^2} 
+ \Delta\right]\Phi =0, \label{WaveEq}
\end{equation}
where $c_s$ is the sound speed. It is a constant 
in space and time for such a medium. 
This equation has Lorentz invariance, that is, if we 
replace the speed of light by the speed of sound, it retains
the form shown above in the new space-time co-ordinates    
obtained after Lorentz-transforming to a frame moving at a
constant speed less than the sound speed. Just as the light field 
{\em in vacuo} is a proper relativistic field, sound is a 
``relativistic'' field.\footnote{More properly, we should 
term this form of Lorentz invariance 
{\em pseudorelativistic} invariance. We will 
however use for simplicity ``relativistic'' as a generic term 
if no confusion can arise therefrom.}
The Lorentz invariance can be made more manifest by writing
equation (\ref{WaveEq}) in the form 
$ \Box \Phi\equiv 
{\sf \eta}^{\mu\nu}\partial_\mu \partial_\nu \Phi = 0 $, where
$\eta^{\mu\nu} = $\,diag($-1,1,1,1)$ is the (contravariant) 
flat space-time metric (we choose throughout the signature 
of the metric as specified here), determining the fundamental
light-cone structure of Minkowski space;\cite{minkowski}     
we employ the summation over equal greek indices $\mu,\nu,\cdots.$ 
Assuming, then, the sound speed $c_s = c_s ({\bm x},t)$
to be local in space and time, and employing the curved space-time version 
of the 3+1D Laplacian $\Box$,\cite{MTW} one can write down the sound wave 
equation in an {\em inhomogeneous medium} in the 
generally covariant form\cite{unruh,visser}  
\begin{equation}
 \frac{1}{\sqrt{\sf -g}} \partial_\mu 
(\sqrt{\sf -g}{\sf g}^{\mu\nu}\partial_\nu \Phi) =0 . 
\label{curvedspaceWaveEq}
\end{equation}   
Here, ${\sf g} = {\rm det} [{\sf g}_{\mu\nu}]$ is the 
determinant of the (covariant) metric tensor. It is to be
emphasized at this point that because the space and time
derivatives $\partial_\mu$ are covariantly transforming objects in 
(\ref{curvedspaceWaveEq}), the primary object in 
the condensed-matter identification of space-time metrics 
via the wave equation (\ref{curvedspaceWaveEq}) is the 
contravariant metric tensor 
${\sf g}^{\mu\nu}$.\cite{GrishaUniverse} In the condensed-matter
understanding of analogue gravity, the quantities  ${\sf g}^{\mu\nu}$ 
are {\em material-dependent} coefficients. They occur 
in a dispersion relation of the 
form ${\sf g}^{\mu\nu}k_\mu k_\nu=0$, where $k_\mu = (\omega/c_s, {\bm k})$ 
is the covariant wave vector, with $\hbar {\bm k}$ the ordinary spatial 
momentum (or quasi-momentum in a crystal).    

The contravariant tensor components 
${\sf g}^{\mu\nu}$, for a perfect, irrotational liquid turn  
out to be\cite{unruh,visser,trautman}
\begin{equation}
{\sf g}^{\mu\nu}= \frac{1}{A_c c_s^2}  \left( \begin{array}{cc} -1 & -{\bm v} 
\\ -{\bm v} & \, c_s^2 {\bm 1} - {\bm v} \otimes {\bm v} 
\end{array} \right),
\end{equation}
where $\bm 1$ is the unit matrix and $A_c$ a  space and time dependent
function, to be determined from the proper equations of motion for the
sound field (see below). 
Inverting this expression according to 
$\textsf{g}^{\beta\nu}\textsf{g}_{\nu \alpha}= \delta^{\beta}{}_\alpha$,  
to obtain the covariant metric ${\sf g}_{\mu\nu}$, 
the fundamental tensor of distance reads 
\begin{equation}
{\sf g}_{\mu\nu}= A_c \left( \begin{array}{cc} -(c_s^2-{\bm v}^2) & -{\bm v} 
\\ -{\bm v} & {\bm 1}
\end{array} \right), \label{gdown} 
\end{equation}
where the line element is $ds^2= {\sf g}_{\mu\nu} dx^\mu dx^\nu$.
This form of the metric has been derived 
by Unruh for an irrotational perfect
fluid described by Euler and continuity equations;\cite{unruh}  
its properties were later on explored in more detail 
in particular by Visser.\cite{visser}
I also mention that an earlier derivation of Unruh's form of the metric 
exists, from a somewhat different perspective;  
it was performed by Trautman.\cite{trautman}

The conformal factor $A_c$ in (\ref{gdown}) depends on the spatial 
dimension of the fluid. It may be unambiguously 
determined by considering the action of the velocity potential fluctuations
above an inhomogeneous background,  
identifying this action with the action of a minimally 
coupled scalar field in $D+1$-dimensional space-time:\cite{PRL,PRD}  
\begin{eqnarray}
S  & = & \int d^{D+1}x 
\frac{1}{2g} \left[ -\left(\frac\partial{\partial t} \Phi -{\bm v} 
\cdot \nabla \Phi\right)^2 + c_s^2  (\nabla \Phi)^2 \right] \nonumber\\
& \equiv & \frac12 \int d^{D+1}x 
\sqrt{-{\sf g}} {\sf g}^{\mu\nu} 
\partial_\mu \Phi \partial_\nu \Phi \,, \label{action}
\end{eqnarray}
where it is assumed 
that the compressibility $1/g$  of the (barotropic) 
fluid, $g = d(\ln m\rho )/dp$, where $p$ is the pressure 
and $\rho$ the density of the fluid, is a constant.
Using the above canonical identification, it may 
easily be shown that the conformal factor is given by
$A_c = \left({c_s}/{g}\right)^{4-D}$. It is mentioned here that 
the case of one
spatial dimension ($D=1$) is special, in that the so-called conformal
invariance in two space-time dimensions implies 
that the classical equations of motion are invariant (take the same form)  
for any space and time dependent choice of the conformal factor $A_c$. 

The line element $ds^2 = {\sf g}_{\mu\nu} dx^\mu dx^\nu$ 
gives us the distances travelled by the phonons 
in an effective space-time world in which 
the scalar field $\Phi$ ``lives''.
In particular, quasiclassical (large momentum) phonons will follow 
{\em light}-like, that is, here, {\em sound}-like geodesics in that 
space-time, according to $ds^2=0$. 
Particularly noteworthy is the simple fact that the
constant time slices obtained by setting 
$dt=0$ in the line element are conformally flat, i.e.
the quasiparticle world looks on constant time slices 
like the ordinary (Newtonian)
lab space, with a simple Euclidean metric in the case of Cartesian spatial 
co-ordinates we display.\footnote{This fact implies that metrics with 
nonvanishing spatial curvature, like the Kerr metric, 
are not amenable within this simple effective metric scheme, which 
starts from Euler and continuity equations; see for a discussion Ref.\,\cite{MattAJP}.}
All the intrinsic curvature of the effective 
space-time is therefore 
encoded in the metric tensor elements ${\sf g}_{00}$ and
${\sf g}_{0i}$. Together with Matt Visser, I described 
this curvature and its properties for an isotachic fluid 
(i.e., having a speed of sound independent of space and time).\cite{AnnPhys}  
Using that phonons move on geodesics, 
we discovered in\cite{PRLMatt} the phenomenon that a vortex acts on 
quasiclassical phonons as an effective gravitational lens.
In Ref.\,\cite{warp}, using the fact that there 
is curvature in any spatially inhomogeneous flow 
(that is, a flow which is not a simple 
superposition of translational motion and rigid body 
rotation), we have shown 
that there exists a sonic analogue of the ``warp-drive'' 
in general relativity
permitting superluminal, i.e., here ``superphononic'' motion.\cite{warpdrive} 
The point made by us is that in the acoustic 
curved space-times we consider, there
is no violation of any condition on the positivity of energy necessary, 
which is in marked contrast to the original warp drives, where 
local energy densities by necessity must be negative to permit
superluminal travel.\cite{Olum} This is due to the fact 
that the Einstein equations, relating curvature and energy-momentum 
content of all fields other than gravitational fields, as already 
mentioned in the above do not need to be imposed in the analogy.

It is important to recognize that the form (\ref{curvedspaceWaveEq}) 
of the wave equation is valid generally (with a possible additional 
scalar potential term). That is, a generally covariant 
curved space-time wave equation can be formulated not just 
for the velocity perturbation (sound) potential in an irrotational 
Euler fluid, for which we have introduced the effective metric concept.
If the spectrum of excitation (in the local rest frame) 
is linear, $\omega = c_{\rm prop} k$, where $c_{\rm prop}$ is the
propagation speed of {\em some} collective excitation, the 
statement that an effective space-time metric exists is 
true, provided we only consider wave perturbations of 
a single scalar field which constitutes the fixed classical background. 
The argument to reach this conclusion is as follows. 

Given that the action density $\cal L$
is a functional of $\phi$ and its space-time derivatives 
$\partial_\mu \phi$, i.e. ${\cal L}= {\cal L} [\phi, \partial_\mu \phi]$, 
we expand the action to quadratic order in the fluctuations
around some stationary classical background solution 
$\phi_0$ of the Euler-Lagrange equations. 
For any Lagrangian of the specified form, the wave equation 
for perturbations $\delta \phi \equiv \Phi$ above the background 
$\phi_0$ is\cite{BLV}
\begin{equation}
\partial_\mu 
\left(\left.\frac{\partial^2 {\cal L}}{\partial(\partial_\mu\phi) 
\partial(\partial_\mu\phi)} \right|_{\phi=\phi_0}
\partial_\nu \Phi\right)
-\left.\left( \frac{\partial^2 {\cal L}}{\partial\phi \partial\phi }
-\partial_\mu \left\{
\frac{\partial^2 {\cal L}}{\partial(\partial_\mu\phi) 
\partial\phi}\right\}\right) \right|_{\phi=\phi_0} \Phi=0.
\label{General}
\end{equation} 
The above equation in covariant notation reads
\begin{equation}
 \frac{1}{\sqrt{\sf -g}} \partial_\mu 
(\sqrt{\sf -g}{\sf g}^{\mu\nu}\partial_\nu \Phi) 
- V_{\rm eff}(\phi_0) \Phi =0 , 
\end{equation}
where $V_{\rm eff}(\phi_0)$ 
is a background-field dependent effective potential 
(equal to the second term in round brackets in equation (\ref{General})
divided by $ \sqrt{-g}$). The potential $V_{\rm eff}$ may, for example, 
in the simplest case contain an effective mass 
of the scalar field, such that the wave equation 
becomes Klein-Gordon like, $\Box \Phi = -m^2 c_{\rm prop}^4 \Phi$.

The effective metric coefficients are,  
up to an (again dimension dependent)  
conformal factor given by: 
\begin{equation}
g_{\mu\nu} (\phi_0) \propto
\left.\frac{\partial^2 {\cal L}}{\partial(\partial_\mu\phi) 
\partial(\partial_\mu\phi)} \right|_{\phi=\phi_0}.
\end{equation} 
The concept of an effective space-time metric therefore 
applies to every system having a single scalar 
wave equation of second order in both space and time derivatives, 
corresponding to perturbations propagating on a 
fixed classical background, where this background itself determines the 
metric coefficients.

\section{The effective metric in Bose-Einstein condensates}
We assumed in Eq.\,(\ref{action}) 
that the compressibility $1/g$  
is a constant.
This entails that the (barotropic) equation of state reads 
$p = \frac12 g \rho^2$. We then have, in terms 
of the interaction between the particles (atoms) 
constituting the fluid, a contact interaction (pseudo-)potential, 
$V({\bm x}-{\bm x}')= g \delta({\bm x}-{\bm x}')$.  
This is indeed the case for the {\em dilute} atomic gases forming 
a Bose-Einstein condensate. Well below the transition temperature, 
they are described to good accuracy 
by the Gross-Pitaevski\v\i\/
mean field equation for the order 
parameter $\Psi \equiv \langle \hat \Psi \rangle$,\footnote{Observe that 
$\langle \hat \Psi \rangle\neq 0$ breaks particle number
conservation (the global U(1) invariance); for a review 
of the consequences see Ref.\cite{leggettBEC}.} representing the 
expectation value of the quantum field operator $\hat \Psi$:
\begin{equation}
i\hbar\frac{\partial}{\partial t}\Psi ({\bm x},t)
= \left[ -\frac{\hbar^2}{2m} \Delta 
+ V_{\rm trap} ({\bm x},t) + g |\Psi ({\bm x},t)|^2 
\right] \Psi ({\bm x},t) . \label{GPEq}
\end{equation}
The Madelung transformation reads $\Psi = \sqrt \rho \exp [i\phi]$, 
where $\rho$ yields the condensate density and $\phi$ is the 
velocity potential; it allows for an interpretation of quantum theory
in terms of hydrodynamics.\cite{madelung}  
Namely, identifying real and imaginary parts on left- and 
right-hand sides of (\ref{GPEq}), respectively,  gives us the two equations
\begin{eqnarray}
-\hbar \frac{\partial}{\partial t} \phi & = &   
 \frac12 m {\bm v}^2 + V_{\rm trap} + g\rho 
-\frac{\hbar^2}{2m} \frac{\Delta {\sqrt\rho}}{\sqrt\rho}
\equiv \mu + p_Q  , \label{JAEq}\\ & &
\hspace*{-3em}
\frac{\partial}{\partial t}\rho + \nabla \cdot(\rho {\bm v}) =   0.
\label{contEq}
\end{eqnarray}
The first of these equations 
is the Josephson equation for the superfluid phase, 
which corresponds to the Bernoulli equation of classical hydrodynamics, 
where the usual velocity potential of irrotational hydrodynamics 
equals the superfluid phase $\phi$ times $\hbar/m$, 
such that ${\bm v} = \hbar\nabla\phi/m$.
The latter equation implies that the flow is irrotational save
on singular lines, around which the {\em wave function phase} 
$\phi$ is defined only modulo 
$2\pi$. Therefore, circulation 
is quantized,\cite{onsagerstathydro} and these singular lines
are the center lines of quantized vortices.
The isothermal chemical potential $\mu $ (which we have chosen 
to incorporate the kinetic energy term $\frac12 m {\bm v}^2$), 
is augmented by the ``quantum pressure term'' 
$p_Q\equiv -\frac{\hbar^2}{2m} {(\Delta {\sqrt\rho})}/{\sqrt\rho}$, 
which is the one genuine quantum term in (\ref{JAEq}), because 
$p_Q\propto \hbar^2$ 
(observe that the first order in $\hbar$ 
may be incorporated into the velocity potential).
The second equation (\ref{contEq}) 
is the continuity equation for conservation of 
particle number, i.e., atom number in the superfluid gas.  
The dynamics of the weakly interacting, dilute ensemble of atoms is thus
that of a perfect Euler fluid with quantized circulation of singular 
vortex lines, which are the only vortical excitations in a superfluid. 
The fact that it is an Euler fluid is true 
 save for regions in which the density rapidly varies
and the quantum pressure term $p_Q$ becomes relevant, which happens 
on scales of order the coherence length $\xi_0 = \hbar/\sqrt{2gm\rho_0}$, 
where $\rho_0$ is a constant asymptotic density far away from the 
density-depleted (or possibly density-enhanced) region. The quantum pressure becomes 
relevant in the depleted-density cores of quantized vortices, or at the low-density 
boundaries of the system, and is negligible outside these domains 
of rapidly varying or low density. 

The whole armoury 
of space-time metric description of excitations,
 explained in the last section, which is 
based on the Euler and continuity equations, is then valid
for phonon-like excitations of a Bose-Einstein condensate, with 
the space-time metric (\ref{gdown}), as long as we are outside the 
core of quantized vortices,\footnote{A treatment of sound 
wave propagation in the presence of vorticity, and the corrections to 
Eq.\,(\ref{curvedspaceWaveEq}) arising therefrom, may be found in 
Ref.\,\cite{MichaelandMatt}.} and far enough from the boundaries
of the condensate.  

\section[Nonuniqueness of the quasiparticle content]{Nonuniqueness of the quasiparticle content of a 
Bose-Einstein condensate}
Quasiparticles are the fundamental entities used to describe
an interacting condensed matter system in a particle picture, that is, 
in a suitable Fock space. 
On the microscopic level, if the elementary constituents
interact by two-body forces, we are given a
second quantized Hamiltonian operator of the form 
$\hat H=\sum_{\bm k} \epsilon_{\bm k} \hat a_{\bm k}^\dagger \hat a_{\bm k} +
V_{{\bm k}{\bm k}'} \hat a^\dagger_{{\bm k}'} \hat a_{\bm k}^\dagger \hat
a_{\bm k} \hat a_{{\bm k}'}$, where
$V_{{\bm k}{\bm k}'}$ are the matrix elements for two-particle 
interaction in a plane wave basis, and $\epsilon_{\bm k} $ 
are the bare single particle energies of the ``elementary'' 
bosons or fermions, which are created by the operators 
$\hat a^\dagger_{\bm k}$ from the proper particle vacuum.    
One then employs a unitary, i.e. 
operator-algebra-conserving Bogoliubov 
transformation\cite{Bogoliubov} to another set of quasiparticle operators 
$\hat {b}_{\bm k}$, 
$\hat {b}^\dagger_{\bm k}$,
see Eq.\,(\ref{Bogotrafo1}) below. This    
gives the Hamiltonian the reduced diagonal form   
$\hat H_{\rm red} = \sum_{\bm k} \omega ({\bm k}) {\hat {b}}_{\bm k}^\dagger {\hat {b}}_{\bm k} + {\rm O}(\hat {b}^3)$, where 
${\rm O}({\hat b}^3)$ represents
additional terms of higher order than quadratic, which are supposed to be
small compared to the leading diagonalized part of the Hamiltonian, for the 
picture of noninteracting quasiparticles to make sense. 
These quasiparticles possess a (possibly spatially anisotropic) 
dispersion relation $\omega  = \omega ({\bm k})$ which is linear for 
various important classes 
of quasiparticles, $\omega(k) \propto k$. Among these classes 
of collective excitations are, e.g., phonons, antiferromagnetic 
magnons,\cite{Kittel} or the excitations around the gap nodes of the $p$-wave superfluid 
$^3\!$He-A.\cite{GrishaUniverse} 

Phonons are the small momentum quanta of the sound field in solids or fluids, with 
$\omega (k) = c_s k$ for a medium at rest. 
Their classical equation of motion in a 
perfect fluid is the generally 
covariant wave equation (\ref{curvedspaceWaveEq}), with the effective space-time 
metric (\ref{gdown}). 
In the simplest case, for a homogeneous medium with space
and time independent density, and a constant speed of sound $c_s$,   
the quantized velocity potential of the phonons reads\cite{llstat2}
\begin{equation}
\hat\Phi ({\bm x},t) =  \sum_{{\bm k}}\sqrt{\frac{{\hbar c_s}}{2{V}\rho_0 k}}
\left[{\hat{b}}_{\bm k}e^{-ic_s k t + i{\bm k}\cdot {\bm x} } + 
{\hat{b}^\dagger}_{\bm k} e^{ic_s k t - i{\bm k}\cdot {\bm x} }\right],
\label{homo}
\end{equation}
where $V$ is the quantization volume of the system, and $\rho_0$ the constant 
background density. That is, the quantum excitation 
field in a homogeneous medium may generally be decomposed into plane waves, 
with the appropriate frequencies as a function of momentum 
stemming from the dispersion relation, here $\omega = c_s k$. 
Note, in particular, that for this
spatially and temporally homogeneous case, the statement that we observe  
positive frequency (energy) with respect to the laboratory time interval 
$dt$ is unique, that is, it can be made independent 
of time and space. In an inhomogeneous fluid, this is (generally) 
no longer the case, 
and the notion of an excitation having positive energy may depend 
on where the detector is located in the fluid, if it is at rest
relative to the fluid or moves, and what its
natural time interval is. The latter may be different 
from that of the laboratory, due to the particular way the detector
couples to the fluid, see section \ref{GibbonsHawking}. 

\subsection{Operator basis dependence of quasiparticle content}
The number of particles assigned to the quantum field $\hat \Phi$ is 
unique with respect to a certain given state $|\Theta\rangle $
of the quantum field, $n_{\bm k}({\Theta}) = 
\langle \Theta | \hat a^\dagger_{\bm k} \hat a_{\bm k}|\Theta\rangle$, 
provided we decompose $\hat \Phi$ into modes associated to the operators
${\hat a}_{\bm k}$ and their Hermitian conjugates.
The number of particles
is, in particular, zero for the {\em vacuum state with respect to the 
field operator $\hat a_{\bm k}$}, 
defined by $\hat a_{\bm k}|0\rangle = 0$. However, it need
not be zero with respect to another set of quasiparticle
operators $\hat {b}_{\bm k}$, which has a {\em different} vacuum
$|\bar 0\rangle$. 

To demonstrate the basis dependence of quasiparticle number, we use 
a general Bogoliubov transformation for bosons, the class of 
(quasi-)particles we consider, of the general 
form 
\begin{eqnarray}
{\hat a}_{\bm i} & = & \sum_{\bm k} \alpha_{{\bm i} {\bm k}} 
\hat {b}_{\bm k} +\beta_{{\bm i} {\bm k}} 
\hat {b}_{\bm k}^\dagger \,, 
\label{Bogotrafo1}
\end{eqnarray}
where $\bm k$ represents a set of quantum numbers, not 
necessarily the plane waves used in Eq.\,(\ref{homo}).
The coefficients in the Bogoliubov
transformation must fulfill certain conditions for the transformation
to be unitary, i.e., to preserve the bosonic commutation relations 
for the new operators, $[\hat b_{\bm i}, \hat b^\dagger_{\bm k}]= 
\delta_{{\bm i}{\bm k}},\, [\hat b_{\bm i}, \hat b_{\bm k}]=0,\,
[\hat b^\dagger_{\bm i}, \hat b^\dagger_{\bm k}]=0$. As a consequence of this 
defining unitary character, the 
following conditions on the transformation coefficients must hold:
\begin{eqnarray}
\sum_{\bm k} \alpha_{{\bm i} {\bm k}}  \alpha^*_{{\bm j} {\bm k}}  
-\beta_{{\bm i} {\bm k}}\beta^*_{{\bm j} {\bm k}} 
=  \delta_{{\bm i}{\bm j}}, 
\qquad 
\sum_{\bm k} \alpha_{{\bm i} {\bm k}} \beta_{{\bm j} {\bm k}} 
-\beta_{{\bm i} {\bm k}} \alpha_{{\bm j} {\bm k}} =  0. \label{Bogocond}
\end{eqnarray}
By using the transformation (\ref{Bogotrafo1}),  
it is straightforwardly shown that the number of $\hat a_{\bm k}$ 
particles in $|\bar 0\rangle$ is given by 
$\langle \bar 0 | {\hat a}^\dagger_{\bm k} {\hat a}_{\bm k}|\bar 0 
\rangle =\sum_{\bm k'} |\beta_{{\bm k'} {\bm k}}|^2$: The old operator 
${\hat a}_{\bm k}$ does not annihilate the new vacuum $|\bar 0 \rangle$
(and vice versa), 
and what looks empty in one quasiparticle vacuum may be full of 
quasiparticles in another.
Related to this (formal) operator-basis dependence is the fact 
that the actually {\em detected} number of 
particles is strongly {\em observer} dependent, 
as opposed to the formally defined quantity $n_{\bm k}(\Theta)$, 
which refers to one particular quasiparticle state $|\Theta\rangle$.  
The detected number of particles depends, in particular,
on how the detector actually couples to the field $\hat \Phi$ whose
quanta it measures. 
Various couplings of the detector, for example to different 
powers of the fluid density, will 
influence the quasiparticle basis in which the detector measures, and 
thus the quasiparticle number detected. 

Now, the salient point is that 
because the phonon field $\hat \Phi$ is a {\em relativistic} quantum field,   
we will be able to map the observer dependence just described, which is 
general and holds for any sort of quasiparticles,  
to the observer dependence 
experienced by proper relativistic quantum fields 
in curved or flat space-time.\cite{unruh76,Gibbons} 
The observer 
dependence is of kinematical origin, i.e., it originates in the fact 
that the relativistic quantum field propagates in a space-time of 
Lorentzian signature, cf. the discussion 
of the Hawking radiation analogue phenomenon in Ref.\,\cite{MattPRL}, and 
is  therefore fully within the capabilities of our proposed analogy.  

\subsection{Scaling ansatz in expanding Bose-Einstein condensates}
To model the quasiparticle analogue of expanding universes, 
we will make use of expanding Bose-Einstein condensates, which are  
produced by reducing the trapping potential strength, i.e.
the harmonic trapping frequency with time, or by increasing the 
interaction coupling constant. 
Firstly, it is thus appropriate 
to describe the evolution of density and velocity distribution in the
expanding gas by discussing 
the so-called scaling procedure established in Refs.\,\cite{KaganPRA,CastinDumPRL96}. 
The scaling procedure introduces a set of generally three scaling variables, 
$b_i=b_i(t)$, which are a function of time only. 
Using these scaling variables, one writes for the (Cartesian) co-ordinate 
vector components $x_{bi} = x_i /b_i $; for the scaled co-ordinate 
vector we use the shorthand notation 
${\bm x}_b \equiv \sum_i {\bm e}_i x_i/b_i $. 
It may then be shown that the evolution of the Bose-Einstein-condensed
gas cloud is described, starting 
from the initial density and velocity potential 
distributions $ \rho = \rho_{\rm init} ({\bm x},t=0)$, 
$\phi=\phi_{\rm init} ({\bm x},t=0)$,  
by the following density and velocity distributions 
(from here on we take $\hbar = m =1$):\cite{KaganPRA,CastinDumPRL96} 
\begin{eqnarray}
\rho({\bm x},t) & \Rightarrow & 
\frac{\tilde{\rho}_{\rm init}({\bm x}_b)}{\mathcal{V}},\label{ScalDensity}\\
{\bm v}= \nabla \phi({\bm x},t) & \quad \Rightarrow\quad & 
{\bm v}= \nabla\phi = \sum_i \frac{\dot{b}_{i}}{b_{i}}x_{i}
+\nabla\tilde{\phi}({\bm x}_{b},t)\,.
\label{ScalVelocity}
\end{eqnarray}
This is true provided the scaling parameters $b_i$ fulfill 
the equations\cite{KaganPRL,PRA}
\begin{equation}
\ddot{b}_{i}+\omega_{i}^{2}(t)b_{i}=
\frac{g(t)}{g(0)}\frac{\omega_{0i}^{2}}
{{\mathcal{V}} b_{i}},
\label{genbEq}
\end{equation}
where $\omega_{0i}$ are the initial trapping frequencies 
and the dimensionless ``scaling volume'' reads $\mathcal{V}=\prod_i b_i$.
The dots are time derivatives with respect to laboratory time $t$ here
and in what follows.\footnote{Note that by special convention 
we generally do not sum over
latin indices $i,j,\cdots$, but only over greek indices
$\mu,\nu,\cdots$.} 
We have taken into account in the above equation 
that the particle interaction can be varied in lab time, 
$g=g(t)$, by means of a suitable Feshbach 
resonance;\cite{feshbach,feshbachII} $g(0)$ is the initial coupling constant.
We designate ``scaling basis'' quantities 
with a tilde. For example, the (stationary) 
initial density distribution 
$\rho_{\rm init}({\bm x})$ gives us 
$\tilde \rho_{\rm init} ({\bm x}_b) $ if we replace
${\bm x} \rightarrow {\bm x}_b$. 
The scaling evolution is exact in a Thomas-Fermi approximation  
which neglects the quantum pressure term $p_Q$ 
and the kinetic energy $\frac12 m {\bm v}^2$. 
Since Bose-Einstein 
condensates were experimentally created, the scaling solution 
has routinely been employed to interpret the time-of-flight 
pictures with which they are visualized as well as analyzed.\cite{Anglin} 

We define a ``scaling time'' variable by 
\begin{equation}
\frac{d\tau_s}{dt}=\frac{g(t)/g(0)}{\mathcal{V}} , 
\label{deftau}
\end{equation}
and the $\tau_s$ dependent {\em scaling functions} $F_i$ by 
\begin{equation}
F_{i}(\tau_s)=\frac{\mathcal{V}}{b_{i}^{2}} \frac{g(0)}{g(\tau_s)}
= \frac{1}{b_{i}^{2}}\frac{dt}{d\tau_s}.
\label{defF}
\end{equation} 
In terms of these quantities, the effective second order action for the 
scaling basis fluctuations of the phase of the superfluid order parameter 
$\delta \tilde \phi ({\bm x}_b, \tau_s)
\equiv \Phi ({\bm x}_b, \tau_s)$ 
takes on the particularly simple diagonal form\cite{PRA} 
\begin{equation}
\bar{S}^{(2)}=\int d\tau_s d^3{x_b}\frac{1}{2g(0)}\left[
-\left(\frac{\partial}{\partial\tau_s}\delta\tilde{\phi}\right)^{2}
+ 
\tilde c^2 F_{i}(\nabla_{bi}
\delta\tilde{\phi})^{2}\right], \label{kappaaction}
\end{equation}
where the scaling speed of sound 
${\tilde c}=\sqrt{g(0) \tilde \rho_{\rm init} ({\bm x}_b)}$. 
Because of the fact that this action does not mix spatial and
temporal derivatives, the resulting 
line element {\em in the scaling variables},  
according to the identification displayed in (\ref{action}) 
is diagonal (does not possess ${\sf g}_{0i}$ terms), and reads 
\begin{equation}
ds^{2} = {\sf g}_{\mu\nu} dx^\mu dx^\nu = \frac{\tilde c}{g(0)}
\sqrt{F_{x}F_{y}F_{z}} \left[-{\tilde c}^{2}d\tau_s^{2}
+F_{i}^{-1}dx_{bi}^{2}\right].\label{blineelement}
\end{equation}
We now consider for simplicity the isotropic case, $b_i \equiv b$, 
which implies 
\begin{equation} 
F_i \equiv F = b^{D-2}\, \frac {g(0)}{g} .
\end{equation}
The generalization of the mode expansion in Eq.\,(\ref{homo}) 
to inhomogeneous expanding Bose-Einstein condensates 
then takes, in this isotropic case, the form\cite{PRA}
\begin{eqnarray}
\hat \Phi ({\bm x}_b, \tau_s)
& = & \sum_{n}\sqrt{\frac{{g(0)}}{2\tilde{V}\epsilon_n}}\phi_{n}({\bm x}_{b})\left[{\hat{b}}_{n}\chi_{n}(\tau_s)+{\hat{b}}_{n}^{\dagger}\chi_{n}^{*}(\tau_s)
\right].\label{3DSol}
\end{eqnarray}
The functions $\phi_{n}({\bm x}_{b})$ are the stationary solutions of the 
Gross-Pita\-evski\v\i\/ equations for excitations above the initial ground state, 
designated by the (set of) quantum numbers $n$ 
with initial energies $\epsilon_n$. The initial Thomas-Fermi 
quantization volume is
$\tilde V$; in the hard-walled cubic box limit the modes
are plane waves, $\phi_n({\bm x_b}) \rightarrow \exp [i{\bm k}\cdot {\bm x}_b] 
$. 

The temporal mode functions $\chi_{n}(\tau_s)$ satisfy 
the second order ordinary differential equation\cite{PRA}
\begin{equation}
\frac{d^{2}}{d\tau_s^{2}}\chi_{n}+F(\tau_s)\epsilon_{n}^{2}\chi_{n}=0.\label{chin}\end{equation}
The case when $F$  is a constant (unity) is particular.  
In this case, the quantum state of the quasiparticle excitations 
remains unchanged for increasing $\tau_s$, and a given initial 
quasiparticle vacuum, {\em in the scaling basis} with 
the associated quasiparticle operators defined by (\ref{3DSol}), 
remains empty forever. That is, no quasiparticles are created 
in that basis, although the superfluid may be in a highly nonstationary 
motional state, obtained by changing the trapping $\omega=\omega(t)$ 
rapidly with time. However, a detector which 
measures in a quasiparticle basis different from the scaling basis, 
for example due to its particular coupling to the superfluid,  
may still detect that quasiparticles are ``created''. We will 
come back to this possibility in section \ref{GibbonsHawking} below, when 
we discuss the purely choice-of-observer related phenomenon
of a thermal state in a quasiparticle basis belonging to one
particular space-time, the de Sitter space-time of cosmology.

\subsection{``Cosmological'' quasiparticle production}
\label{Cosmological} 
Consider now the general case that the scaling function 
$F (\tau_s)$ is a function of scaling time $\tau_s$.  
The fact that $F$ depends on time implies that the statement ``the 
excitation is of positive frequency'' (a particle) 
or of ``negative frequency'' (antiparticle) for a 
given propagating wave cannot be held up for all 
times $\tau_s$. This {\em frequency mixing} implies that quasiparticles
are  created from the quasiparticle vacuum, because an initially empty
scaling basis vacuum
state does not remain empty during the evolution of the system, i.e.
initially $\hat b_n |0 (\tau_s =0) \rangle = 0 $, but at a later
stage $\hat b_n |0 (\tau_s)\rangle \neq 0 $. 

The fact that annihilation and creation operator parts
of the initial vacuum are mixed, as a consequence of (\ref{Bogotrafo1}), is 
physically due to the fact that quasiparticles are scattered within the
course of time at (time dependent) effective potentials. 
Physically, the $\tau_s$ dependence of $F$ furnishes such an 
effective potential, 
at which excitations are scattered from 
negative to positive frequency and vice versa: 
The equation (\ref{chin}) is formally equivalent to scattering of a 
non-relativistic particle with energy $\epsilon_{n}$ by a potential 
in $\tau_s$ space, 
\begin{equation} 
V(\tau_s)= \epsilon_{n}^{2}(1-F(\tau_s)).
\end{equation}
At large $\tau_s$,  the WKB scattering solution of (\ref{chin})
therefore reads:\cite{PRA}  
\begin{equation}
\chi_{n} =\frac{1}{F^{{1}/{4}}}
\left(\alpha_{n}\exp\left[-i\epsilon_{n}\int_{-\infty}^{\tau_s} {d\tau_s'}
\sqrt{F(\tau_s')}\right]
+\beta_{n}\exp\left[i\epsilon_{n}\int_{-\infty}^{\tau_s} {d\tau_s'}\sqrt{F(\tau_s')}\right]\right),
\label{chinEqs}
\end{equation}
where the scattering amplitudes are related via the 
particle flux conservation condition $|\alpha_{n}|^{2}- |\beta_{n}|^{2}=1$.
The quantity $N_{n}=|\beta_{n}|^{2}$ can be interpreted as the number of 
{\em scaling basis} quasiparticles created from the initially empty 
scaling vacuum,  due to the time dependent scattering 
of excitations moving in the nonstationary condensate.

In the WKB approximation, the amplitudes 
$\alpha_n$ and $\beta_n$ are connected in a simple way:
\begin{equation}
\beta_{n}=\exp\left[-\frac{\epsilon_{n}}{2T_{0}}\right]\alpha_{n},
\label{alphaT0beta}
\end{equation}
where the inverse temperature $1/T_0$ is given by the integral 
\begin{equation}
\frac{1}{T_{0}}
={\Im}\left[\int_{\cal C}\sqrt F \, d\tau_s\right], \label{defT0}
\end{equation}
and $\cal C$ is the contour in the complex $\tau_s$-plane enclosing the
closest to the real axis singular point of the function 
$\sqrt{F(\tau_s)}$.\cite{ll} 
This gives us the number of particles created in the mode $n$, 
using the flux conservation condition $|\alpha_{n}|^{2}- |\beta_{n}|^{2}=1$, 
\begin{equation}
N_n=|\beta_{n}|^{2}=\frac{1}{\exp[\epsilon_{n}/T_{0}]-1}.
\end{equation}
The distribution of the created quasiparticles 
follows a thermal bosonic distribution (Planck spectrum), at a
temperature $T_0$. The adiabatic evolution of trapped gases hence  
leads to {}``cosmological'' quasiparticle creation with thermal occupation 
numbers in the scaling basis.
The temperature $T_{0}$ occurring in the Planck distribution above 
depends on the details of the scaling evolution, i.e., on   
the specific superfluid dynamics imposed by the solution of 
Eqs.\,(\ref{genbEq}), that is, $T_0$ is a functional of the temporal evolution 
$\omega_i = \omega_i(t)$.\cite{BLVFRW,PRA} 

We use the term ``cosmological'' in context with
the plain condensed-matter fact that quasiparticles are created in 
the scaling basis. We now justify this by comparing our effective 
Bose-Einstein condensate metric 
(\ref{blineelement}) to line elements which constitute 
cosmological solutions of the Einstein equations.
For example, let $d\tau_s=dt$ by properly adjusting 
$g=g(t)$, thus choosing the coupling constant's time evolution to be given by 
$g(t)= g(0){\mathcal V}(t) $, cf. Eq.\,(\ref{deftau}). We then obtain that 
(\ref{blineelement}) equals (up to the conformal factor 
${\tilde c}/(g(0) {\mathcal V}$)) 
an anisotropic version of the spatially flat Friedmann-Robertson-Walker (FRW)
universe:\cite{Weinberg} 
\begin{equation}
ds^{2}=\frac{\tilde c}{g(0){\mathcal V}}
\left[-\tilde c^{2}dt^{2}+ \sum_i b_i^{2}d x_{bi}^{2}
\right].
\label{blineelementiso}
\end{equation}
In the standard spatially isotropic form of the FRW metric, 
all $b_i = b$ are equal.
 
To obtain the exact equivalence to a 
spatially flat FRW metric, 
we have to assume in addition that 
$\tilde c$ is spatially independent,  
which is fulfilled close to the center of the gas cloud, where the
parabolic density profile is approximately flat, and $\tilde c$ is 
essentially a constant. 
In the spatially isotropic case, 
\begin{equation}
H = \frac{\dot b}b \label{defH}
\end{equation}
is the Hubble ``constant'', which obviously is a  
constant only if the gas expands exponentially
in laboratory time, $b \propto \exp [Ht]$, just as we 
need exponentially rapid expansion for a constant Hubble 
parameter in inflationary cosmological models.\cite{inflation,RMPinflaton} 

We therefore come to the remarkable conclusion 
that the co-ordinate scaling factor $b$ of the 
Bose-Einstein condensate quasiparticle universe, occurring 
in the equations of motion of a nonrelativistic 
condensed matter system, has exactly the same meaning, in the
quasiparticle world, as the cosmological 
scale factor of the Universe proper.

\section{Gibbons-Hawking effect in de Sitter space-time}
\label{GibbonsHawking}
We now treat the case that the scaling factors $F_i$ in (\ref{defF})
are constants, i.e., 
do not depend on scaling time. 
Therefore, following Eqs. (\ref{3DSol}) and (\ref{chin}), 
no scaling basis quasiparticles are created through negative and 
positive frequency mixing. The superfluid can still 
be in highly nonstationary motion, though: The time evolution is 
according to (\ref{defF})
prescribed by 
\begin{equation}
b_i^2(t) = C_i \frac{g(0)}{g(t)} {\mathcal{V}(t)} \quad 
\Longleftrightarrow \quad
\frac{b_i}{\prod_{k\neq i} b_k} = C_i \frac{g(0)}{g(t)}, 
\end{equation}
cf. Eq. (\ref{defF}), where the $C_i$ are constants.
However, although the superfluid is in motion, no dissipation through 
intrinsic quasiparticle creation takes place, 
because there exists the  Fock space ``scaling'' 
basis, in which no quasiparticles 
are created from the scaling vacuum, and the energy of 
that particular superfluid vacuum is conserved. A particular instance is
the isotropic 2D case, $b_1=b_2 = b$, where for constant $g(t) = g(0)$
the condition on $F$ being constant is fulfilled.\footnote{Cf. 
Ref.\,\cite{PitaRosch}, where the fact of superfluid vacuum energy 
conservation is explained from a different (SO(2,1) symmetry) perspective, 
and Ref.\,\cite{DalibardBreathing} 
for a quality factor measurement of breathing (monopole) oscillations 
in a cylindrical geometry.}  

In de Sitter space-time space is empty and flat (that is, 
the constant time slices are Euclidean space), and all 
of the curvature of space-time is encoded in a nonvanishing 
cosmological constant $\Lambda$.  For obvious reasons, the 
de Sitter space-time is very popular 
in the quantum field theoretical treatment of cosmological 
theories,\cite{BirrellDavies,HawkingEllis} because it highlights the 
crucial cosmological role the energy density of all conceivable 
quantum fields taken together might play: The vacuum energy density 
may constitute the dominant effective source term for space-time curvature 
in the Einstein equations. 

The Gibbons-Hawking effect for geodesic observers\cite{Gibbons} 
in such a de Sitter space-time is the curved space-time 
 analogue of the
Unruh-Davies effect.\cite{BirrellDavies,unruh76} The latter consists 
in the fact that a constantly accelerated detector moving in the flat
(purportedly ``empty'') space-time vacuum, responds as if it 
were placed in a thermal bath of (quasi-)particles with temperature 
proportional to its acceleration. 
{Observer-related} phenomena are at the heart of quantum 
field theory on nontrivial, and generally curved, space-time backgrounds.
They tell us, in particular, that the particle content of a given quantum field 
depends on the (motional) state of the detection apparatus, 
which is verifying that there are particles by its ``clicks''.
More technically speaking, the dependence of the particle
content of quantum fields in curved space-time is rooted in  
the non-uniqueness of canonical field quantization in 
Riemannian spaces.\cite{fulling} It is of fundamental importance
to make these observer-dependent effects measurable, because such 
a measurement constitutes, {\it inter alia}, 
a  consistency check for an all-important 
concept of standard quantum field theory, namely 
that the quantization of 
a given field is carried out on a {\em fixed} space-time background.
 
That the experimental verification of the observer dependence 
is exceedingly difficult with light becomes readily apparent if we 
calculate the Unruh-Davies temperature:
The result is that it equals 
$ T_{\rm Unruh} = 
[\hbar /(2\pi k_ {\rm B} c_{\rm L})]  a = 4\, {\rm K} \times a[10^{20} g_{\oplus}] \,,$ 
where $a$ is the acceleration of the detector in Minkowski space 
($g_{\oplus}$ is the gravity acceleration on the surface of the Earth), and 
$c_{\rm L}$ the speed of light. The huge accelerations needed to obtain
measurable values of $ T_{\rm Unruh}$ 
make it obvious that an observation of the 
effect with light (photons) is decidedly 
a less than trivial undertaking. Although proposals for a 
measurement with ultraintense short pulses of electromagnetic radiation have 
been put forward in, e.g., Refs.\cite{Yablonovitch,chenUnruh}, 
it is less than obvious how the thermal spectra associated to the
effect, which still furnish tiny contributions to the total 
energy, should be discernible from the background 
dominated by the ultraintense lasers used to 
create large accelerations of the elementary particles contained 
in the plasma. 

\subsection{An isotropic de Sitter universe in a harmonic trap}
We first discuss the simplest case of an isotropically expanding gas 
in a harmonic trap. We will see that this case also serves 
a pedagogical purpose, because it forces us to 
distinguish between purely observer-related phenomena with thermal spectrum, 
which by definition all have $F_i(\tau_s)\equiv1$, and 
``cosmological'' particle creation with a thermal spectrum, 
as defined in section \ref{Cosmological}, for which the 
scaling functions $F_i(\tau_s)$ in (\ref{defF}) 
depend explicitly on scaling time. 
 
One can create de Sitter universes in an 
expanding gas by letting it expand exponentially, $b(t) \propto \exp[Ht]$
where $H$ is the Hubble constant of cosmological expansion, equalling 
$H=\Lambda/D$ for the de 
Sitter universe discussed, where $D$ is the spatial dimension.  
The de Sitter metric in its standard form, using as its time co-ordinate  
the ``cosmological'' time $\tau_c$, reads\cite{Weinberg}
\begin{equation}
ds^{2}=\frac{c_0}{g(0){\mathcal V}}
\left[-c^2_0 d\tau_c^{2}+ e^{2H\tau_c} d {\bm x}_{b}^{2}
\right],\label{deSittercomoving}
\end{equation}
where $c_0 \equiv \tilde c({\bm 0})$ is the central scaling (i.e., initial)
speed of sound. 
In the present case of exponential expansion, 
the ``cosmological'' time interval occurring in the
metric above is equal to both the laboratory and the scaling time interval, 
$d\tau_c =d\tau_s = dt$; cf. Eq.\,(\ref{blineelementiso}) 
with $b_i=b = \exp[H\tau_c]$.

However, using the isotropic harmonic expansion  
setup, one has to face the difficulty
that in order to give the FRW 
metric (\ref{blineelementiso})
near the center of the trap (i.e., close to ${\bm x}_b =0$)  
the de Sitter form, one has to increase exponentially 
the interaction with laboratory time: 
Because of (\ref{defH}), 
$g(t) \propto b^D \propto \exp[DHt] = \exp[c_0 \Lambda t]$. 
Though the central density also decreases exponentially (like
$b^{-D/2}$), the exponential increase of the coupling ``constant'' incurs
strong three-body recombination losses,\cite{PetrPRL96} whose total rate 
(in the dilute gas case and in three spatial dimensions)  
is proportional to $g^4 \rho^2 \propto b^{9} = \exp [9  c_0 \Lambda t]$. 
Therefore, within a short
time of order $1/H$, the Bose-Einstein-condensed
 gas of interacting single atoms
will simply 
no longer be in existence, because the atoms rapidly form bound states. 
Such an experiment will leave no time to measure
phenomena which depend on the fact that an equilibrium is established; 
in particular, the thermal equilibrium for the occupation numbers 
in the de Sitter quasiparticle basis will not be established on such 
a short time scale. 

Even more importantly, though one obtains indeed a thermal spectrum
in the de Sitter quasiparticle basis corresponding to the metric
(\ref{deSittercomoving}), $F(\tau_s) = F(t) = 1/b^2 = \exp[-2Ht]$
depends on time, i.e., it is not a constant in the quasiparticle basis 
corresponding to the mode functions 
$\chi_n \propto \exp[-i\epsilon_n \tau_s ]$.  
The thermal spectrum obtained therefore is, by our physical definition,
the  thermal ``cosmological'' quasiparticle creation discussed in the
previous section. It is {\em not} what we want to observe in our 
effective de Sitter space-time, namely the purely choice-of-observer 
related phenomenon Gibbons-Hawking effect, for which no actual quasiparticle
``creation'' in the scaling basis should take place.

\subsection{The 1+1D de Sitter universe in a cigar-shaped cloud}
\begin{center}
\begin{figure}[b]
\psfrag{zH}{$z_{\rm H}$}
\psfrag{-zH}{$-z_{\rm H}$}
\psfrag{b}{\large $b\propto t$}
\psfrag{bperp}{\large $b_\perp\propto \sqrt t$}
\psfrag{Atomic}{Atomic}
\psfrag{Quantum Dot}{Quantum Dot}
\vspace*{1em}
\centerline{\epsfig{file=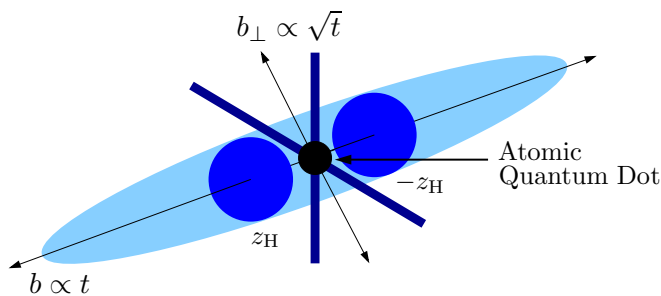,width=0.66\textwidth}}
\caption{\label{ellipse} Expansion of a cigar-shaped Bose-Einstein 
condensate, with scale factor $b$ linear in lab time along the axial, 
and with $\sqrt t$ in the radial direction, scaled by $b_\perp$. 
The stationary horizon surfaces 
are located at $\pm z_{\rm H}$, respectively. 
The thick dark lines represent lasers creating an optical potential 
well in the center of the harmonic trap, which hosts the Atomic Quantum Dot, 
cf. Fig.\,\ref{AQD}.}
\end{figure}
\end{center}

To circumvent the problem that the interaction coupling 
needs to be increased exponentially with time to obtain a 
de Sitter universe in 2+1 or 3+1 isotropic space-time dimensions
in an isotropic harmonic trap, I 
and Petr Fedichev have developed the model of a 1+1D de Sitter universe.
This 1+1D toy model can be realized in a 
strongly anisotropic, cigar-shaped Bose-Einstein condensate,\cite{PRL,PRD} 
cf. Fig.\,\ref{ellipse}. In particular, 
in the proposed experimental setup, no time variation of the 
coupling constant at all is necessary, 
which is thus a true ``constant'' also in time.

\vspace*{-1em}
\vspace*{-1.5em}

The analysis of the excitation modes in a strongly anisotropic, elongated 
Bose-Einstein condensate is based on the adiabatic separation ansatz\cite{Zaremba} 
\begin{equation}
\Phi (r,z,t) = \sum_n \phi_n(r) \chi_n(z,t), \label{sepansatz}
\end{equation}
where $\phi_n(r)$ is the radial wavefunction characterized by the 
quantum number $n$ (only zero angular momentum modes are considered here). 
The above ansatz incorporates the fact that for strongly elongated 
traps, i.e., traps for which $\omega_z \ll \omega_\perp$,
 the dynamics of the condensate
motion separates into a fast radial motion and a slow axial motion, which 
are essentially independent. 
The $\chi_n(z,t)$ are the mode functions for travelling wave solutions in the 
$z$ direction (plane waves for a condensate at rest read 
$\chi_n \propto \exp [- i \epsilon_{n,k} t + kz]$). The radial motion 
is assumed to be ``stiff'' such that the radial part is effectively 
time independent, because the radial time scale for adjustment of the 
density distribution after a perturbation is much less than the axial 
oscillation time scales of interest.  
The ansatz (\ref{sepansatz}) works independent 
from the ratio of healing length and 
radial size of the superfluid cigar.  In the limit that the healing length 
is much less than the radial size, Thomas-Fermi wave functions are used, 
in the opposite limit, a Gaussian ansatz for the radial part 
of the wave function $\phi_n(r)$ is appropriate.  

The squared oscillation spectrum of the cigar-shaped condensate cloud
reads $\epsilon^2_{n,k} 
=c_0^2 k^2  + 2\omega_{\perp}^2 n (n+1)$, 
where $c_0 =\sqrt{\mu/2} $.\cite{Zaremba}  
A de Sitter space-time is then obtained from the effective action for the
phase fluctuations of the phonon ($n=0$) modes near the 
center of the trap. 
The corresponding 1+1D action is obtained after integrating out the 
transverse, strongly confined directions, and reads:\cite{PRL,PRD}
\begin{eqnarray}
S_0  &= & \int dt  dz \frac{\pi b_\perp^2 R_{\perp}^2}{2g}
\left[ -\left(\frac{\partial}{\partial t} \chi_0 -{v}_z \partial_z\right)^2 
+ \frac{c^2_0}{b_{\perp}^2b} (\partial_z \chi_0)^2 
\right] \nonumber\\
& \equiv & \frac12 \int d^{D+1}x 
\sqrt{-{\sf g}} {\sf g}^{\mu\nu} 
\partial_\mu \chi_0 \partial_\nu \chi_0 \,.
\label{1daction}
\end{eqnarray}
The scaling parameters $b$ in the axial ($\omega_z$)
and $b_\perp$ in the perpendicular ($\omega_\perp$) directions
are functions of time, such that the action fulfills the 
identification with the action of a scalar field 
minimally coupled to gravity, 
as indicated in the second line 
above, where the metric coefficients are 
the ${\sf g}_{\mu\nu}$ in Eq.\,(\ref{gdown}). 

The identification of the above phase-fluctuations action with the action 
of a scalar minimally coupled to gravity
works, that is, the two actions in the first and second lines  
of (\ref{1daction}) are consistent, if we impose the 
consistency condition that  
\begin{eqnarray}
\frac{ \pi b_\perp^2 R_\perp^2 }g Z^2 
& = & \frac{b_\perp\sqrt {b}}{c_0} \nonumber\\
\quad \Longleftrightarrow \quad
\frac{b_\perp}{{\sqrt b}} & = & 8 \sqrt{\frac{\pi}2} 
\frac1{Z^2} \sqrt{\rho_m a_s^3} 
\left(\frac{\omega_\perp}\mu\right)^2 \equiv B 
= {\rm const}.,
\label{cond1}
\end{eqnarray} 
where $Z$ is a renormalization factor according 
to $\chi_0 = Z \bar \chi_0 $, with $\bar \chi_0$ the renormalized
wave function, and $\rho_m$ the initial central density. 
The factor $Z$ does not influence the (classical) 
equation of motion $\delta S/\delta \chi_0 = 0 $ (it simply drops out), 
but does influence the response of a detector. We will come 
back to this point in section \ref{deSitterdetect} below, when we 
discuss the explicit dependence on $Z$ of the  
equilibration time scale for the stationary detector state considered, 
see Eq.\,(\ref{tauequil}).

In Refs.\cite{PRL,PRD}, we were using an 
alternative form of the 1+1D de Sitter metric (\ref{deSittercomoving}).  
This alternative form 
is the one used in the original Gibbons-Hawking paper.\cite{Gibbons} 
It reads (we leave out the conformal factor)
\begin{equation}
d s^2  = -c_0^2 \left(1-\Lambda  z^2 \right)  
d {\tau}^2 + \left({1-\Lambda z^2 }\right)^{-1} 
d z^2\,. 
\label{deSitterlineelement}
\end{equation}
The time interval $d\tau$ in the above metric is {not} the 
``cosmological'' time interval $d\tau_c$ in the 
version of the metric displayed in (\ref{deSittercomoving}). 
The two metrics may be transformed into each other using the following 
co-ordinate
transformations:
\begin{eqnarray}
\exp[-2\sqrt \Lambda c_0 \tau] & = & \exp[-2\sqrt \Lambda c_0 \tau_c] 
-\Lambda z_b^2 ,
\nonumber\\ 
z &= & z_b \exp[\Lambda c_0 \tau_c]  , 
\label{coordtrafo} 
\end{eqnarray}
giving us $\tau = \tau (\tau_c,z_b)$, $z = z (\tau_c,z_b)$, which transforms 
(\ref{deSitterlineelement})  into (\ref{deSittercomoving}).
The advantage of the form (\ref{deSitterlineelement}) is that it is plain 
in this form that the de Sitter space-time has an event horizon, located
in our 1+1D case at the constant values $z= z_{\rm H}= \pm \Lambda^{-1/2}$. 

The quantity $\Lambda z^2$ in the de Sitter metric (\ref{deSitterlineelement})
must be
independent of time. This leads us to the requirement 
\begin{equation}
\sqrt\Lambda z = \frac{v_z}{c(t)} = 
\frac{\dot b b_\perp }{\sqrt b c_0 }z =
\frac{B \dot b}{c_0} z
\end{equation}
relating the dynamical 
parameters of our problem to each other 
($c(t)$ is the instantaneous 
sound velocity at the center of the cloud). These relations imply  
that $\dot b =$ const., and thus that $b\propto t$.  Then, 
the cosmological constant $\Lambda$ becomes independent of time, 
as is necessary for an analogue de Sitter space-time to be established. 

The experimental procedure now is determined to be as 
follows: Prepare a Thomas-Fermi (i.e., sufficiently large), 
strongly anisotropically trapped, cigar-shaped  
condensate. 
Let it expand in the axial direction linear
in lab time by changing the trapping according to the scaling
equations (\ref{genbEq}), such that $b \propto t$, {\em and} 
simultaneously expand in the perpendicular direction with 
the square root of lab time, $b_\perp \propto \sqrt t$, such that 
$B= b_\perp/\sqrt b$ in (\ref{cond1}) is a constant.
 Then, a detector ``tuned'' to the de Sitter space-time 
(\ref{deSitterlineelement}), i.e., which works in the de Sitter
quasiparticle basis, will measure a thermal quasiparticle spectrum, 
with the de Sitter temperature:\cite{PRL,PRD,Gibbons}  
\begin{eqnarray}
T_{\rm dS} 
=   \frac{c_0}{2\pi } \sqrt{\Lambda} = \frac B{2\pi}  {\dot b}  .
\label{TdS}
\end{eqnarray}
The fact that a thermal spectrum is obtained can be directly derived 
from the equations of superfluid hydrodynamics, as expounded in Refs.\,\cite{PRL,PRD}. 
That is, it is not simply postulated due to the (kinematical) 
analogy with quantum field theory in de Sitter space-time, 
but is a result  of the quantized hydrodynamic 
equations determining the evolution of the
quasiparticle content of the sound field in the de Sitter basis. 

The relation between de Sitter time $\tau$ and the laboratory time
is fixed by $d\tau = dt/({\sqrt{b} b_\perp})= dt/(b(t) B) = dt/(\dot b t B)$.  
[Note that $d\tau$ and the scaling time interval $d\tau_s$ defined
in Eq.\,(\ref{deftau})  differ; 
$d\tau_s = dt /B^2 b^2 = d\tau /Bb $.]
The  transformation law 
between $t$ and the de Sitter time $\tau$ 
(on a constant $z$ detector, such that $d\tilde t = dt $), is 
given by
\begin{equation}
\frac{t}{t_0} = \exp[B \dot b \tau], \label{trafo} 
\end{equation} 
where the unit of lab time $t_0 \sim \omega_\parallel^{-1}$ 
is set by the initial conditions 
for the scaling variables $b$ and $b_\perp$.   

It is important to recognize that 
an effective exponential ``acceleration'' of the oscillation frequencies, 
either because of the exponential dependence of
laboratory time on scaling time, represented by (\ref{trafo}),  
or coming from the WKB approximation for the 
scattering amplitudes, and the exponentially small 
mixing of positive and negative frequency parts resulting 
therefrom, Eq.\,(\ref{alphaT0beta}), 
is sufficient for the thermal spectrum to be obtained. 
In particular, though we have discussed a space-time which possesses a 
horizon, no pre-existing horizons in the parent space-time
are necessary {\it per se} for thermal occupation number distributions to be 
established, as has been pointed out in Ref.\,\cite{HuRaval}. 
An explicit example is the Unruh-Davies effect, where
this parent, global space-time is simply Minkowski space.

\subsection{Detecting the thermal de Sitter spectrum}
\label{deSitterdetect} 
To detect the Gibbons-Hawking effect in de Sitter 
space-time, one has to set up a detector which measures 
frequencies in units of the inverse de Sitter time 
$\tau$, rather than in units of the inverse laboratory time $t$; 
this corresponds to measuring in the proper de Sitter 
quasiparticle vacuum, where, in particular, 
positive and negative frequency are defined with respect to $\tau$.
Only then does one detect quasiparticles which are defined with respect to
the de Sitter quasiparticle basis,  
and refer to a vacuum corresponding to exactly that space-time.

\begin{center}
\begin{figure}[t]
\psfrag{U}{\large $U$}
\psfrag{Omega}{\large $\Omega$}
\psfrag{Vopt}{\large $V_{\rm opt}$}
\psfrag{l}{\large $l$}
\psfrag{alpha}{\large $|\alpha \rangle$}
\psfrag{beta}{\large $|\beta \rangle$}
\psfrag{Delta}{\large $\Delta$}
\vspace*{-8em}
\centerline{\epsfig{file=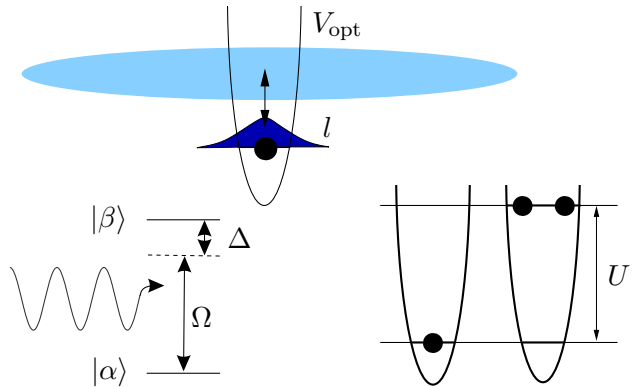,width=0.66\textwidth}}
\caption{\label{AQD} 
Level scheme of the ``Atomic Quantum Dot'', which is embedded in 
the superfluid cigar, and created by an optical well for atoms
of a hyperfine species different from that of the condensate. 
Double occupation of the dot is prevented by a 
collisional blockade mechanism.}
\end{figure}
\end{center}

In Refs.\,\cite{PRL,PRD}, 
I and Petr Fedichev have 
provided such a
detector. We have shown that a ``de Sitter basis'' detector is 
realized by an ``Atomic Quantum Dot'' (AQD) 
(for a detailed exposition of AQD properties 
cf. Ref.\,\cite{AQDAlessio}). 
The AQD can be implemented in a Bose gas of atoms possessing two hyperfine
ground states $\alpha$ and $\beta$; the level scheme is  represented
in Fig.\,\ref{AQD}. 
The atoms in the state $\alpha$ represent the expanding 
Bose-Einstein condensate, and are used to model the  expanding  
de Sitter universe. 
The AQD itself is formed by trapping atoms in the state $\beta$ in a 
tightly confining optical potential $V_{\rm opt}$ created by 
a laser at the center of the cloud. 
The interaction of atoms in the two internal levels is
described by a set of coupling parameters $g_{cd} = 
4\pi a_{cd}$ ($c,d =\{\alpha,\beta\}$),
where $a_{cd}$ are the $s$-wave scattering
lengths characterizing short-range intra- and inter-species collisions;
$g_{\alpha\alpha}\equiv g$, $a_{\alpha\alpha} \equiv a_s$, and 
$g_{\alpha\beta}\equiv \bar g$.
The on-site repulsion between the atoms $\beta$ 
in the dot is given by the energy level spacing 
$U\sim g_{\beta \beta}/l^{3}$ between states with a occupation difference of
one $\beta$ atom,
where $l$ is the characteristic size of the ground state wavefunction
of atoms $\beta$ localized in $V_{\rm opt}$. 
We consider the so-called collisional
blockade limit of large $U>0$, where only one atom of type
$\beta$ can be trapped in the dot. This 
limit assumes that $U$ is much larger
than all other relevant frequency scales in the dynamics of both
the AQD and the expanding superfluid, and corresponds to a 
large ``Coulomb blockade gap'' in 
electronic quantum dots.\cite{Manninen}  
As a result of these assumptions, 
the collective co-ordinate of the AQD is modeled by a 
pseudo-spin-$1/2$ degree of freedom $\bar\eta$, with 
spin-up/spin-down state corresponding to 
occupation of the AQD by a single atom or no atom 
in the hyperfine state $\beta$. A Rabi laser of frequency $\Omega$, with 
a detuning $\Delta$ from resonance between the two hyperfine levels 
$\alpha$ and $\beta$, couples atoms of the hyperfine species $\alpha$, 
constituting the expanding cigar-shaped superfluid, 
into the AQD. 
\vspace*{-2.5em}

The detector Lagrangian reads (Ref.\,\cite{PRD}, cf. the 
Hamiltonian formulation in Ref.\,\cite{PRL}):    
\begin{eqnarray}
L_{\rm AQD} & = &  i 
\left(\frac{d}{dt} {{\bar \eta}^*} \right) 
\bar \eta 
- \Omega \sqrt{\rho_0(0,t) l^3}
\left(\bar \eta + \bar \eta^*\right) 
\nonumber  \\& & 
-\left[-\Delta+(\bar g-g)\rho_0 (0,t)
+\bar g \delta \rho 
+ \frac d{dt} \delta\phi
\right] \bar \eta^* \bar \eta\,. \label{Lbareta}
\end{eqnarray}
The fact that the detector has the de Sitter basis as its ``natural''
quasiparticle basis, and therefore measures in de Sitter time, 
is due to the fact that the term linear in the detector co-ordinate
$\bar \eta$  in the Lagrangian couples in a certain manner 
to the superfluid cigar in which it is embedded.
Namely, the laser with frequency $\Omega$, causing transitions 
between the two hyperfine levels $\alpha$ and $\beta$, couples to the  
{\em square root} of the central mean-field particle density. 
This particular coupling (represented by the second term 
in the first line of the above Lagrangian) is what we need, because 
$\sqrt{\rho_{0}(0,t)} = \sqrt{\rho_m}/bB= \sqrt{\rho_m}/(\dot b B t) 
=  \sqrt{\rho_m} d\tau/dt$. The fact that the coupling coefficient 
is proportional to $d\tau/dt$ 
is required to establish that the detector can work as a de Sitter detector,
because it transforms the detector equations, which are obviously 
{\it a priori} in laboratory time, into equations in de Sitter time; 
see the equations (43) for the temporal evolution of the detector 
level occupations below.

Adjusting the detuning $\Delta$ properly, 
such that $\Delta (t) =(\bar g-g)\rho_{0}(0,t)=
(\bar g-g)\rho_m/(\dot b^2 B^2 t^2)$, the 
first and the second term in the square brackets in the 
second line of (\ref{Lbareta}) cancel. One then obtains a simple set of 
coupled equations for the occupation 
amplitudes of the state $\psi
= \psi_\beta |\beta\rangle +\psi_\alpha |\alpha\rangle $ of the AQD: 
\begin{equation}
i\frac{d \psi_\beta}{d\tau}=\frac{\omega_0}2 \psi_\alpha + \delta V \psi_\beta,\,\,\,\,\qquad 
i\frac{d \psi_\alpha}{d\tau}=\frac{\omega_0}2 \psi_\beta,\label{deSitterEvol}
\end{equation}
where $d\tau$ is the {\em de Sitter time interval.} 
Were it not for the density oscillations in the cigar, 
represented by the potential $\delta V$, the 
above equations (\ref{deSitterEvol}) would represent a simple 
two-level system, with frequency splitting 
$\omega_0=2\Omega \sqrt{\rho_m l^ 3}$. 
The density oscillations contained in the perturbation operator  
$\delta V (\tau) = \left(\bar g-g \right)B b(\tau) \delta \rho(\tau)$
cause transitions between the two undisturbed {\em detector eigenstates}
 $|\pm\rangle= (|\alpha\rangle \pm |\beta\rangle)/\sqrt 2$
of the two-level system, which are separated by the energy $\omega_0$. 
The density perturbations in the expanding host superfluid  
lead to a damping of the Rabi 
oscillations with frequency $\omega_0$ between these two states. This  
constitutes the effect of the de Sitter thermal bath 
to be observed, where the damping happens on the time scale 
displayed  in (\ref{tauequil}) below. 

The response of the detector, that is, the transition rates between 
the detector states, can 
be calculated by evaluating a 
response function\cite{BirrellDavies} which makes
use of the expectation value of the product of two 
$b(\tau) \hat\rho(\tau)$ operators.
The  probability per unit time for 
excitation ($P_+$, transition from $|+\rangle$ to $|-\rangle $) 
and de-excitation ($P_-$, transition from $|-\rangle$ to $|+\rangle $) of the
detector takes the form:\cite{PRD,unruh76}
\begin{eqnarray}
\frac{dP_{\pm}}{d\tau} &= &\lim_{{\cal T}\rightarrow\infty}
\frac{1}{\cal T}\int^{\cal T}\!\!\!\int^{\cal T} 
\! d\tau d\tau^{\prime}\langle\delta \hat V(\tau)\delta \hat V(\tau^{\prime})
\rangle e^{\mp i\omega_{0}(\tau-\tau^{\prime})}
\nonumber\\
&= &\lim_{{\cal T}\rightarrow\infty}
\frac{B^2 \left(\bar g-g \right)^2}{\cal T}\int^{\cal T}\!\!\!\int^{\cal T} 
\! d\tau d\tau^{\prime}\langle b(\tau)\delta \hat \rho(\tau) b(\tau^\prime)\delta \hat \rho(\tau^{\prime})
\rangle e^{\mp i\omega_{0}(\tau-\tau^{\prime})}.
\label{detectorresponse}
\end{eqnarray}
The second-quantized solution of the hydrodynamic
equations for the density fluctuations above the superfluid ground state  
in the expanding cigar-shaped Bose-Einstein condensate reads\cite{PRL,PRD} 
\begin{eqnarray}
\delta\hat\rho =  \sum_k
i\sqrt{\frac{\epsilon_{0k}}{4 \pi R_\perp^2 R_\parallel g}} 
\frac  \partial {\partial t}  
\left( 
\hat a_{k} \exp \! \left[
-i\int^t\! \frac{dt' \epsilon_{0k}}{Bb^2}+ ik z_b \right]
\right) \!
+ {\rm H.c.} \label{deltarho} 
\end{eqnarray}
Using this solution, and inserting into (\ref{detectorresponse}), 
we have shown\cite{PRL,PRD} that at late times $\tau$ the 
transition probabilities per unit de Sitter, i.e., per unit detector 
time satisfy detailed balance conditions. They correspond to 
{\em thermodynamic equilibrium} at the temperature $T_{\rm dS}$ 
displayed in (\ref{TdS}):
\begin{equation}
\frac{dP_{+}/d\tau}{dP_{-}/d\tau} = \frac{n_{\rm B}}{1+n_{\rm B}}, 
\label{occup}
\end{equation} 
where the Bose-Planck distribution function takes the form
familiar from 
thermostatistics:
\begin{equation}
n_{\rm B}=\frac1{\exp[\omega_{0}/T_{\rm dS}]-1]}.
\end{equation} 
The frequency $\omega_0\propto \Omega $ can be varied by 
changing the undressed Rabi frequency $\Omega$, 
varying the intensity of the Rabi laser. The detector 
thus has a changeable and therefore {\em tunable} frequency 
standard, which can be adjusted to scan the above distribution function for 
a given de Sitter temperature $T_{\rm dS}$.

From the relation (\ref{occup}), 
we come to the remarkable conclusion that a properly designed 
detector can ``see'' a thermal equilibrium distribution
in its quasiparticle basis, though it is
embedded in a highly nonstationary system {with respect
to the laboratory frame}. 
The rapidly expanding Bose-Einstein condensate represents this
highly nonstationary system, which
hosts the de Sitter quasiparticle detector AQD.  
I stress here again that (\ref{occup}) is an exact {\em result} obtained 
by quantizing hydrodynamic fluctuations in a nonstationary 
superfluid, and not just concluded from a mere comparison of the phonon 
dynamics in our expanding superfluid with 
the quantum field theory of photons in de Sitter space-time.

The equilibration time scale of the detector, and thus the time scale
on which the Rabi oscillations between the detector states are damped
out, is set by the detector frequency standard (the level spacing) 
 $\omega_0$, and by the 
renormalization factor $Z$:\cite{PRD}  
\begin{equation}
\tau_{\rm equil} = 
Z^{-2} \omega_0^{-1} 
\propto ({\rho_m a_s^3})^{-1/2} 
\left(\mu/{\omega_\perp}\right)^2 \omega_0^{-1} .
\label{tauequil}
\end{equation}
The renormalization factor $Z$ contained in 
(\ref{cond1}) determines the equilibration rapidity because
it physically expresses the strength of detector-field coupling. 
It is related to the initial diluteness parameter 
$D_p(0) \equiv ({\rho_m a_s^3})^{1/2}$ of the
Bose-Einstein condensate and to the ratio $\mu/\omega_\perp$, which determines
inasmuch the system is effectively one-dimensional, 
$Z^2 \propto D_p(0) (\omega_\perp/\mu)^2$. 
To obtain sufficiently fast equilibration, the condensate thus 
has to be initially not too dilute as well as close to the quasi-1D
r\'egime, for which the transverse harmonic oscillator energy scale 
is of order the energy per particle, $\mu \sim \omega_\perp$. These two conditions 
have another important implication. 
The ratio of the instantaneous coherence length 
$\xi_c (t) = (8\pi \rho_0(0,t) a_s)^{-1/2} \propto t $ 
and the location of the horizons $z= z_{\rm H}= \pm \Lambda^{-1/2}$, 
which are {\em stationary} in the present setup, 
has to remain less than unity 
within the equilibration time scale.\footnote{I thank 
C. Zimmermann for a pertinent
question during a talk given by me at T\"ubingen, 
leading to this observation.}  
If this is not the case, the coherence length, which 
plays the role of the analogue Planck scale (which is time
dependent here), exceeds the length 
scale of the horizon at equilibration, 
and the concept of ``relativistic'' phonons
propagating on a fixed curved space-time background with local Lorentz symmetry 
becomes invalid. 
The ratio $\xi_c (t) /z_H$ at the lab equilibration time scale 
$t= t_{\rm equil} = t_0 \exp[2\pi (T_{\rm dS}/\omega_0) 
\left(\mu/{\omega_\perp}\right)^2 D^{-1}_p(0)]$ 
following from the de Sitter equilibration time in Eq.\,(\ref{tauequil}), 
expressed in parameters relevant to the experiment, is given by 
\begin{equation}
\frac{\xi_c (t_{\rm equil})}{z_{\rm H}} = 
\frac{\pi t_0 T^2_{\rm dS}}{\rho_m a_s} 
\exp\left[2\pi \frac{T_{\rm dS}}{\omega_0}  
\left(\frac{\mu}{\omega_\perp}\right)^2 \frac1{D_p(0)}\right].
\end{equation}
We see that this ratio changes exponentially with both the
initial diluteness parameter $D_p(0)$ 
and the quasi-1D parameter $\mu/\omega_\perp$.
In most currently realized Bose-Einstein condensates,\cite{Anglin}  
the diluteness parameter $D_p \sim 10^{-2}$. 
Here, we initially need 
$D_p (0)\sim O(1)$ to have the condition 
${\xi_c (t_{\rm equil})}/{z_{\rm H}}< 1$ fulfilled, 
assuming a reasonably large value of the de Sitter temperature 
$T_{\rm dS}$.  
Though the condensate has to be {\em initially} 
quite dense, it is to be stressed that the central density 
decays like $t^{-2}$ during expansion. Therefore, the 
rate of three-body recombination losses quickly decreases
during the expansion of the gas, and the initially relatively 
dense Bose-Einstein condensate, which would rapidly decay
if left with a $D_p$ close to unity, 
can live sufficiently long, the total rate of three-body 
losses decreasing like $\rho^2_0(0,t) \propto t^{-4}$. 

\section{Summary}
The primary statement to be drawn from the present article is that 
phonons, i.e., low-energy linear-dispersion quasiparticles, moving 
in a spatially and temporally inhomogeneous Bose-Einstein-condensed 
superfluid gas, are equivalent to photons, the quanta of
the electromagnetic field, moving on geodesics in a given curved space-time. 
We have explored the classical as well as the quantum aspects of this
statement. 

On the classical side, the analogy helps to provide us with 
a simple general means to study quasiparticle propagation in an 
inhomogeneous medium in motion. An example of such an application 
is the gravitational lensing effect exerted by a superfluid vortex.\,\cite{PRLMatt} 
On the quantum field theoretical side, we can access within the 
analogy phenomena which are extremely difficult if not impossible 
to access with light. One of the fundamentals of quantum field theory, 
the fact that the particle content of a quantum field depends on the observer, 
can thus be experimentally verified for the first time. 
The basic reason that the phenomena in question are (comparatively) easy 
to simulate in a condensed matter system is that the energy and temperature 
scales, under which they occur, relative to the typical energy scales
of the system, can be changed at will by the experimentalist in a 
very controlled manner.  
More particularly, the temperature of the 
thermal spectrum of phonons to be measured in the 
Gibbons-Hawking effect can be made
relatively large compared to the axial phonon frequencies 
and the actual temperature of the gas itself, by expanding
the condensate cloud rapidly enough. In the condensed 
matter analogue, the typical energy of the quanta produced 
can in principle be even made to approach the relevant 
``Planck'' scale, i.e., the point in the quasiparticle 
energy spectrum where it begins to deviate from being linear.

Finally, on a more adventurous side, one could conceive of 
carrying out experiments in ``experimental'' cosmology, 
as opposed to the currently existing purely ``observational'' cosmology.   
In such an experimental approach to matters cosmological, one would 
try to reproduce under certain specified and, in particular, well-defined 
initial conditions
large-scale features of the cosmos, in the laboratory setting of  
nonstationary, inhomogeneous superfluid gases.



\section*{References}
\begin{thebibliography}{99}
\bibitem{MTW} C.\,W. Misner, K.\,S. Thorne, and J.\,A. 
Wheeler, {\em Gravitation} (Freeman, San Francisco, 1973).
\bibitem{Anglin} J.\,R. Anglin and W. Ketterle,
Nature {\bf 416}, 211 (2002).
\bibitem{Khalatnikov} I.\,M. Khalatnikov, {\em An Introduction to the 
Theory of Superfluidity} (Addison Wesley, Reading, MA, 
1965).
\bibitem{BirrellDavies} N.\,D. Birrell and P.\,C.\,W. Davies,  
{\em Quantum Fields in Curved Space} (Cambridge University Press,  Cambridge, England, 
1984). 
\bibitem{minkowski} H. Minkowski,  
Physik. Zeitschr. {\bf 10}, 104 (1909).
\bibitem{unruh} W.\,G. Unruh, 
Phys. Rev. Lett. {\bf 46}, 1351 (1981).
\bibitem{visser} M. Visser, 
Class. Quantum Grav. {\bf 15}, 1767 (1998).
\bibitem{trautman} A. Trautman, \emph{Comparison of Newtonian
and relativistic theories of spacetime}, in: {}``Perspectives in
Geometry and Relativity'' (Indiana University Press, Bloomington, 1966).
\bibitem{PRL} P.\,O. Fedichev and U.\,R. Fischer, 
Phys. Rev. Lett. {\bf 91}, 240407 (2003). 
\bibitem{PRD} P.\,O. Fedichev and U.\,R. Fischer, 
Phys. Rev. D {\bf 69}, 064021 (2004). 
\bibitem{MattAJP} M. Visser, 
gr-qc/0309072.
\bibitem{AnnPhys} U.\,R. Fischer and M. Visser, 
Ann. Phys. (N.Y.) {\bf 304}, 22 (2003).
 \bibitem{PRLMatt} U.\,R. Fischer and M. Visser, 
Phys. Rev. Lett. {\bf 88}, 110201 (2002). 
\bibitem{warp} U.\,R. Fischer and M. Visser, 
Europhys. Lett. {\bf 62}, 1 (2003). 
\bibitem{warpdrive} M. Alcubierre, 
Class. Quantum Grav. {\bf 11}, L73 (1994).
\bibitem{Olum} K.\,D. Olum, 
Phys. Rev. Lett. {\bf 81}, 3567 (1998).
\bibitem{BLV} C. Barcel\'o, S. Liberati, and M. Visser, 
Class. Quantum Grav. {\bf 18}, 3595 (2001).
\bibitem{leggettBEC} A.\,J. Leggett,  
Rev. Mod. Phys. {\bf 73}, 307 (2001); {\it ibid.} {\bf 75}, 1083(E) (2003).
\bibitem{madelung} E. Madelung, 
Z. Phys. {\bf 40}, 322 (1927).
\bibitem{onsagerstathydro} L. Onsager, 
Nuovo Cimento Suppl. {\bf 6}, 279 (1949). 
\bibitem{MichaelandMatt} S.\,E. Perez Bergliaffa, K. Hibberd,
M. Stone, and M. Visser, 
Physica D {\bf 191}, 121 (2004).  
\bibitem{Bogoliubov} N.\,N. Bogoliubov, {\em Selected Works, Part II: 
Quantum and Statistical Mechanics} (Gordon and Breach, New York, 1991). 
\bibitem{Kittel} C. Kittel, {\em Quantum Theory of Solids}  
(Wiley and Sons, New York, 2nd revised printing 1987).  
\bibitem{GrishaUniverse} G.\,E. Volovik,   
{\em The Universe in a Helium Droplet} (Oxford University Press, Oxford, 2003).
\bibitem{llstat2} E.\,M. Lifshitz and L.\,P. Pitaevski\v\i,  
{\em Statistical Physics, Part 2} (Pergamon Press, New York, 1980).   
\bibitem{unruh76} W.\,G. Unruh, 
Phys. Rev. D {\bf 14}, 870 (1976).
\bibitem{HawkingEllis} S.\,W. Hawking and G.\,F.\,R. Ellis,  
{\em The Large Scale Structure of Space-Time} 
(Cambridge University Press, Cambridge, England, 1973).
\bibitem{Gibbons} G.\,W. Gibbons and S.\,W. Hawking, 
Phys. Rev. D \textbf{15}, 2738 (1977). 
\bibitem{MattPRL} M. Visser, 
Phys. Rev. Lett. {\bf 80}, 3436 (1998).
\bibitem{fulling} S.\,A. Fulling, 
Phys. Rev. D {\bf 7}, 2850 (1973).
\bibitem{Yablonovitch} E. Yablonovitch, 
Phys. Rev. Lett. {\bf 62}, 1742 (1989).
\bibitem{chenUnruh} P. Chen and  T. Tajima,
Phys. Rev. Lett. {\bf 83}, 256 (1999). 
\bibitem{KaganPRA} Yu. Kagan, E.\,L. Surkov, and G.\,V. Shlyapnikov, 
Phys. Rev. A {\bf 54}, R1753 (1996).
\bibitem{CastinDumPRL96} Y. Castin and R. Dum, 
Phys. Rev. Lett. {\bf 77}, 5315 (1996).
\bibitem{KaganPRL} Yu. Kagan, E.\,L. Surkov, and G.\,V. Shlyapnikov,
Phys. Rev. Lett. {\bf 79}, 2604 (1997).
\bibitem{feshbach} S. Inouye {\it et al.}, 
Nature {\bf 392}, 151 (1998). 
\bibitem{feshbachII} A. Marte {\it et al.}, 
Phys. Rev. Lett. {\bf 89}, 283202 (2002).
\bibitem{BLVFRW} C. Barcel\'o, S. Liberati, and M. Visser, Phys. Rev. A {\bf 68}, 
053613 (2003).
\bibitem{PRA} P.\,O. Fedichev and U.\,R. Fischer,
Phys. Rev. A {\bf 69}, 033602 (2004).
\bibitem{ll} L.\,D. Landau, E.\,M. Lifshitz, and L.\,P. Pitaevski\v\i\/, 
{\em Quantum Mechanics: Non-Relativistic Theory} 
(3rd edition, Butterworth-Heinemann, Newton, MA, 1981).
\bibitem{Weinberg} S. Weinberg, {\em Gravitation and Cosmology}    
(Wiley and Sons, New York, 1972).
\bibitem{inflation} A.\,H. Guth, 
Phys. Rev. D \textbf{23}, 347 (1981).
\bibitem{RMPinflaton} J.\,E. Lidsey, A.\,R. Liddle, E.\,W. Kolb, E.\,J. 
Copeland, T. Barreiro, and M. Abney,
Rev. Mod. Phys. {\bf 69}, 373 (1997).  
\bibitem{PitaRosch} L.\,P. Pitaevski\v{\i} and A. Rosch,
Phys. Rev. A {\bf 55}, R853 (1997).
\bibitem{DalibardBreathing} F. Chevy, V. Bretin, P. Rosenbusch, K.\,W. Madison,
and J. Dalibard, 
Phys. Rev. Lett. {\bf 88}, 250402 (2002).
\bibitem{PetrPRL96} P.\,O. Fedichev, M.\,W. Reynolds, and G.\,V. Shlyapnikov,
Phys. Rev. Lett. {\bf 77}, 2921 (1996).
\bibitem{Zaremba} E. Zaremba, 
Phys. Rev. A {\bf 57}, 518 (1998).
\bibitem{HuRaval} B.\,L. Hu and A. Raval, 
Mod. Phys. Lett. A {\bf 11}, 2625 (1996). 
\bibitem{AQDAlessio} A. Recati, P.\,O. Fedichev, 
W. Zwerger, J. von Delft, and P. Zoller, 
cond-mat/0404533.
\bibitem{Manninen} S.\,M. Reimann and M. Manninen,
Rev. Mod. Phys. {\bf 74}, 1283 (2002).  
\end{thebibliography}
\end{document}